\documentclass[final]{ias2}
\usepackage{graphicx} 
\usepackage{multirow}
\usepackage{array} 

\usepackage{hyperref} 

\newcommand\beq{\begin{equation}}
\newcommand\eeq{\end{equation}}
\newcommand\bea{\begin{eqnarray}}
\newcommand\eea{\end{eqnarray}}

\begin{document}
\markboth{Pramana class file for \LaTeX 2e}{Murthy M V N, et. al.}

\title{Anomalous Kolar events revisited: Dark Matter?}

\author[imsc]{M.V.N. Murthy}
\email{murthy@imsc.res.in}

\author[imsc,cmi]{G. Rajasekaran}
\email{graj@imsc.res.in}

\address[imsc]{The Institute of Mathematical Sciences, Chennai 600113, 
India.}

\address[cmi]{Chennai Mathematical Institute, Siruseri 600103, India.}

\begin{abstract}

The possibility of the unexplained Kolar events, recorded in the 70's 
and 80's, being due to the decays of dark matter particles of mass in 
the range of 5--10 GeV is pointed out.

\end{abstract}

\maketitle

\section{Introduction}

Many years ago, in the cosmic ray neutrino experiments 
\cite{krishnaswamy1} and later in the proton decay experiment 
\cite{krishnaswamy2} both at Kolar Gold Fields (KGF) in south India, 
some unusual events, so called Kolar events, were seen. The neutrino 
experiments at KGF were conducted by groups from Tata Institute of 
Fundamental Research, India, Durham University, UK and Osaka City 
University, Japan. They used techniques that were perfected over many 
years for detecting muons with scintillation triggers and Neon Flash 
Tubes for tracking. Seven such detectors were deployed in a long tunnel 
at a depth of 2300 metres in Champion reef mines starting by the end of 
1964. The  Kolar events were multi-track events with some unusual 
features which could not be explained away by any known processes of 
muons or neutrinos. The two sets of Kolar events were interpreted 
\cite{gr1,sarma,gr2}, at that time, as due to the possible decay of a 
new, massive, long-lived particle produced mostly in neutrino or 
antineutrino collisions within the surrounding rock of the mine. We 
however note that searches were also made at the $\nu$- beam experiments 
at CERN \cite{faissner} and at Fermilab \cite{benvenuti} but they led to 
negative results with bounds on cross-sections (and masses) to produce 
such neutral, long-lived particles in neutrino interactions. Thus the 
events were neither confirmed in other experiments nor shown to be 
spurious by any further analyses.

Individually such events could be caused by neutrinos (or antineutrinos) 
interacting with air molecules in the gap between rock wall and the 
detectors. Such events are rare and occur with a probability less than 
one in hundred years. Thus the cause of events numbering close to ten in 
as many years (live time) has remained a source of puzzle since their 
observation.

In this short note we speculate on the possibility of these anomalous 
events being due to the decay of dark matter particles. Darkmatter 
particles are ubiquitous and are present every where with varying 
densities.  This also naturally explains why they were not seen in 
accelerator experiments with neutrino beams \cite{faissner, benvenuti}. 
In section II, we give a brief description of the Kolar events and why 
they were considered anomalous. In section III, we discuss the 
possibility that these events may be caused by the decay of dark matter 
particles and make some remarks about further investigations in this 
direction.

\section{Kolar events}

The Kolar events were recorded over two periods: The first period 
corresponds to the period starting from the end of 1964 (for a review of 
KGF experiments and details of detectors, see Ref.\cite{vsn}). In all 
seven neutrino telescopes, with a geometry that is sensitive to 
horizontal tracks, were installed over a period of two years in a long 
tunnel at a depth of 2300 metres underground. The live time of all 
detectors combined was more than ten years. The first results on Kolar 
events from this period were published in 1975 \cite{krishnaswamy1}.

The second period refers to the experiments set up to look for proton 
decay at 2300 metres depth. Proton decay experiments were done in two 
phases with a live time of 8.41 years and 5.53 years respectively from 
1980-1990. In all about 8 events, encompassing the both periods, were 
found anomalous and these were referred to as Kolar events. 

\begin{figure}[htp]
\centering
\includegraphics[width=1.1\textwidth]{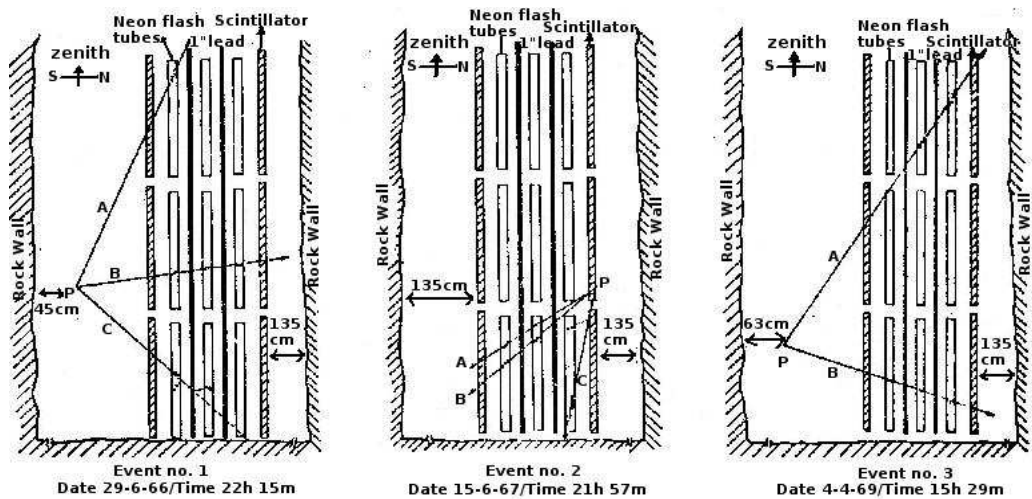} 
\caption{Multi-track (Kolar) events recorded in KGF neutrino-detectors in
the first period. Reproduced from Ref.\cite{vsn} (not to scale).}
\label{fig1}
\end{figure}

A few examples of such events recorded by telescopes 1 and 2  at
2300 metres depth are shown in Fig.\ref{fig1}.

The characteristics of the 5 events reported in 1975 
\cite{krishnaswamy1} are as follows:

\begin{enumerate}

\item In the observed decays, the events consisted of two or more tracks 
with a large opening angle with at least one being a muon as seen from 
the penetrating power.

\item All tracks of an event seemed to originate from a vertex located 
either in air or in the thin detector materials - based on an 
extrapolation of projected angles of tracks. This is the most crucial
fact about these events which renders them anomalous. 

\item The ratio of the number of events containing such tracks to the 
total number of events recorded by the detectors was about 25\%. Such a 
rate can not be explained by direct neutrino or antineutrino 
interactions.

\item On the other hand, in the initial explanations, it was assumed 
that a new particle was produced in the neutrino interaction with the 
rock which eventually decayed in the air. The estimated cross section 
for the production of an assumed new particle multiplied by the 
branching ratio for the observed modes was estimated to be $10^{-37}$ 
cm$^2$ per nucleon, similar to the weak interaction cross-section.

\end{enumerate}

During the second period, each of the 3 events reported in 1986 
\cite{krishnaswamy2} at a depth of 2.3 km at KGF had a penetrating track 
and an associated shower. The details of the events are given in Table 
\ref{tab1}.
\begin{table}[bhp]
\centering
\begin{tabular}{|l|l|l|l|l|}\hline
Event& Penetrating & Shower & Opening & Vertex\\
Number& track(GeV) & (GeV) & angle (deg) & \\
\hline 
1 & $>$ 1.3 & $>$ 2.6 & 32 & air \\
2 & $>$ 0.4 & $>$ 2.5 & 69 & air or rock \\
3 & $>$ 1   & $\ge$ 5 & 41 & inside detector \\\hline
\end{tabular}
\caption{Summary of the 3 events reported in 1986. The track and 
shower energies are given in GeV.}
\label{tab1}
\end{table}

Several theoretical attempts were made 
\cite{gr1,sarma,gr2,derujula,pati} to understand the Kolar events. Both 
sets of events reported in 1975 and 1986 were interpreted as due to the 
decays of an unstable particle, produced in the rock medium by neutrino 
interactions, with a life time approximately given by $10^{-8}$ seconds 
and with a mass in the range $2-5$ GeV. The decay of this new particle, 
in air or in the thin part of the detector, was expected to produce the 
signature as seen in Kolar events. While the events reported in 1975 
were assumed as due to the decays of a charged particle since there were 
three visible charged tracks, the later events were interpreted as due 
to a neutral particle decaying into a muon and possibly an electron, the 
electron producing the shower. However, the Kolar events have so far 
remained an enigmatic puzzle with no conclusive evidence emerging from 
other such detectors around the world. Since the Standard Model is now 
firmly established, any attempt to introduce a new particle with mass in 
GeV range and standard model interaction will be suspect.

\section{Reinterpretation-- decay of dark matter particles}

In the present note, we attempt a reinterpretation of these 8 Kolar 
events as due to the possible decay of a neutral dark matter particle 
(DMP), at rest, of mass around $5-10$ GeV and with a very long life time 
of the order of the life time of the universe, that is $\ge 10^{10}$ 
years . Though the existence of dark matter has been established beyond 
doubt, the nature of DMP is yet to be understood. In particular not much 
attention has been paid to the possibility of DMP decays. A nominal 
lower limit on its life time, $\tau > 10^9$ years\cite{ibarra}, emerges 
from very simple considerations of observed dark matter density at 
present time. A detailed model independent analysis of cosmological 
constraints on decaying dark matter gives a bound on the life time of 
decaying dark matter as $\tau \ge 10^{11}$ years\cite{amigo}. However, a 
combination of stable and decaying dark matter scenario relaxes the 
constraint on the life time and yields $\tau \ge 5\times 10^{9}$ 
years~\cite{ichiki}.

In contrast to the earlier interpretation of the Kolar events, we are 
now disassociating the events from neutrinos interacting in the 
surrounding rock. It had an inherent difficulty of explaining the large 
(25\%) production cross section of the new particle and that difficulty 
disappears with the DMP interpretation now. The DMP is present 
everywhere. Since the DMP's are mostly nonrelativistic, their decays 
must be isotropic. In the Kolar events, the tracks were seen only in one 
hemisphere; it is therefore possible that there were other unobserved 
tracks, particles not going through the detector, that would make the 
decay isotropic. As a result, the earlier estimates of the mass of 
around $2-5$ GeV, using visible energy, must be regarded as a lower 
limit.

Invoking the isotropy of events for DMP decay, it is more likely that 
the mass of DMP will be in the range of 5--10 GeV of which about 2-5 GeV 
is assumed to be deposited in the detectors situated in one hemisphere. 
Furthermore, these unobserved particles in the decay must be charged in 
the events reported in 1975 so that it is consistent with the hypothesis 
of a neutral DMP overall. We note that the CDMSII collaboration 
\cite{cdms} have recently claimed the observation of 3 events in a Si 
detector which are interpreted as due to the nuclear recoil induced by a 
DMP with a most probable mass of 8.2 GeV. This mass is well within the 
range that one would estimate from the Kolar events after accounting for 
isotropy. The announcement of this result, in fact, provided the 
motivation for us to go back and take a re-look at the Kolar events. 
However, some doubt about these events has been cast by the recent 
results from the Large Underground Xenon (LUX) experiment\cite{lux}. No 
final word on CDMSII result has, however, been said yet.

We denote the local number density of DMP in the solar system as $n$. 
Note that this is not the average density in the universe which is much 
smaller than this. The local DMP energy density in the solar system is 
expected to be in the range of GeV/cc \cite{partdata}. If the effective 
volume of the detector chamber sensitive to the decay events of DMP  
is $V$, the mean life of DMP is $\tau$ and the branching 
ratio to the decay into visible modes is $B$, then the rate of decay 
events seen is given by
\begin{equation} R= \frac{nVB}{\tau}. 
\label{eq1} 
\end{equation} 
Furthermore, if we choose $V=10m~\times~10m~\times~10m = 10^9~cc$, $n=1/cc$, 
$B\approx 1$ and $\tau\approx 10^{10}$years, we get a rate $R\approx 
0.1$ decays per year. It is remarkable that such a crude estimate agrees 
roughly with the order of magnitude of the rate of events seen in Kolar.

A few points are in order here:
\begin{itemize}

\item The estimated volume $V=10^9$ cc is probably an underestimate. It 
is more likely to be around $10^{10}$ cc. The experiments were carried 
out at different times with different sized chambers and over a period of 
several years. Hence the volume estimate is at best crude.

\item The number density of DMP $n$ locally is  less than 1/cc 
if we assume the DMP mass in the range of 5--10 GeV. The most recent 
estimates, based on a detailed model of our galaxy including rotation 
curves, give the local DM density to be around $0.39 \pm 
0.03~GeV/cc$~\cite{partdata}. This reduction will be compensated by a 
possible underestimation in the volume under consideration.

\item For simplicity, we have assumed the branching ratio to visible 
modes to be unity.

\item The life time of about $10^{9}--10^{10}$ years, approximately the 
age of the universe, for DMP decay is tantalising. This is well within 
the life time bound based on cosmological constraints with stable and 
unstable dark matter scenario \cite{ichiki}. On the other hand if the 
life time is much more, say about $10^{11}$ years or more, then it may 
be impossible to observe such decays given their density at the present 
epoch.  The present interpretation of Kolar events as due to DMP decay 
will not be valid any more.

\item It is possible that not all Kolar events may be interpreted as DMP 
decays. If we restrict to those with vertex in the air, not in rock or 
inside the detector material, then the observed rate of Kolar events is 
lower and closer to the estimate of R obtained from eq.(\ref{eq1}). This 
is a more likely scenario since the events in rock and detector material 
could be caused by neutrino interactions. The probability of the Kolar 
events being due to atmospheric neutrino interactions in the surrounding 
air is $\le 10^{-3}$ events per year for neutrino energies greater than
5 GeV.

\end{itemize}

One apparent problem with the interpretation of Kolar events as due to 
DMP decay is its non-observation in other detectors. Earlier searches at 
CERN and Fermilab proved negative but they were looking for a 
short-lived particle produced in neutrino interactions at accelerators 
following early theoretical interpretations based on models which are 
now discarded. Since these experiments specifically involved neutrino 
beams interacting with target material inside the detector the negative 
result is easily understood.

It is also unlikely that such events may be identified in neutrino 
detectors such as Super Kamioka (SK) or Sudbury Neutrino Observatory 
(SNO) since there is no (or very little) air gap between the detector 
material (water) and the surrounding rock. As such even if a DMP decays, 
its signature would be submerged in the huge background of neutrino 
events unless the back-to-back geometry can be used to isolate such 
events. Therefore, it may be useful to have a re-look at such of those 
events which conform to the isotropy of all decay products.

On the other hand, it is possible that such anomalous events may be seen 
at MINOS or OPERA, where the detector position is similar to the KGF 
experiment- the detector is placed in a chamber with a large air gap 
between the detector and the rock. However, since the rate is 
approximately 0.1 events a year or less, any non-observation of such 
events in these detectors may still lie within statistical fluctuations. 
Nevertheless the scenario outlined by us in this note should provide 
motivation for such searches at existing detectors or in the proposed 
future underground neutrino detectors like NOVA and INO. The effective 
volume at INO, due to the size of the proposed chamber, is at least 
$10^{11}$ cc. This would immediately increase the rate to 1 event per 
year without compensating for the aperture. 

It is difficult to make clear prediction about such searches without 
using a consistent model for a light, 5-10 GeV, DMP decay. The models 
with heavy DMP decay predict a life time which is much higher ($\ge 
10^{26}~s$) than required for explaining Kolar events \cite{ibarra} 
and therefore are not of much use in the present context.

Finally, we conclude with some general remarks. Independent of the 
estimates given above, the DMP decay hypothesis should be examined more 
closely since all the Kolar events could not have been caused by 
neutrinos or antineutrinos. As such there are no other known sources or 
explanation of these events. The veracity of this claim can only be 
established by new and dedicated experiments. 

If the speculation outlined in this paper is proved correct it solves 
two problems in one stroke-- interpretation of anomalous Kolar events 
and the observation of dark matter particle.

In the same vain and even independent of the Kolar events and their 
interpretation, any large underground detector must be in a position to 
see the decays of an unstable DMP. Therefore, we have one more window 
for searching for DMP provided it decays. Non-observation of the decays 
may be used to set limits on its life time. In fact, the absence of 
spectacular high-energy decay events in past and present large 
underground detectors already rules out life times of the order of 
$10^{10}$ years or less, for heavy DMPs of mass larger than 100 GeV.

Acknowledgements: We thank Pijush Bhattacharjee, Vivek Datar, Shrihari 
Gopalakrishna, Romesh Kaul, N K Mondal, V S Narasimham, Sandip Pakvasa 
and Rahul Sinha for discussions and valuable comments.


\begin{thebibliography}{99}

\bibitem{krishnaswamy1}
M.R.Krishnaswamy et al, Phys. Lett., {\bf 57B},105(1975);
Pramana, {\bf 5},59 (1975).

\bibitem{krishnaswamy2}
M.R.Krishnaswamy et al,  Proc. XXIII Int.Conf. on High Energy Physics,
Berkeley(ed.) S Loken (World Scientific, 1986).

\bibitem{gr1}
G. Rajasekaran and K.V.L. Sarma, Pramana {\bf 5},78 (1975).

\bibitem{sarma}
K.V.L. Sarma and L. Wolfenstein, Phys. Lett. {\bf B61},77 (1976).

\bibitem{gr2}
A.S. Joshipura, G. Rajasekaran, V. Gupta and K.V.L. Sarma, 
Pramana {\bf 33},639 (1989).

\bibitem{faissner}
H. Faissner et al., Phys.Lett., {\bf B60},401 (1976).

\bibitem{benvenuti}
A.C. Benvenuti et al.,Phys. Rev.Lett.{\bf 32},125 (1974); 
ibid,1454(1974); Phys.Rev.Lett.{\bf 35},1486 (1975).

\bibitem{vsn}
V.S. Narasimham, {\it Perspectives in Neutrino Physics}, Proc. Indian
National Science Academy, {\bf 70A},11 (2004).

\bibitem{derujula}
A. de Rujula, H. Georgi and S.L. Glashow, Phys.Rev.Lett.{\bf 35},628 (1975).

\bibitem{pati}
J.C. Pati and A. Salam, Preprint ICTP/75/73, 1975.

\bibitem{ibarra}

M. Garny, A. Ibarra, D. Tran and C. Weniger, J. Cosmology and Astrophysics,
{\bf 1101}, 32 (2011); see also the talk by A Ibarra--
http://kitpc.itp.ac.cn/dsu2011/slides/DSU2011-A.Ibarra.pdf.

\bibitem{amigo}

S. De Lope Amigo, W.M. Cheung, Z. Huang and S. Ng, 
hep-ph:arXiv:0812.4016v2 (2009)

\bibitem{ichiki} 

K. Ichiki, M. Oguri and K. Takahashi, Phys. Rev. Lett. {\bf 93} 
071302(2004).

\bibitem{cdms}
R. Agnese et al, ``Dark matter search results using Silicon detectors
of CDMSII",  hep-ex: arXiv:1304.4279 (2013).

\bibitem{lux}
D.S. Akerib et al, "First results from the LUX dark matter experiment
at the Sanford Underground Research Facility", astro-ph.CO: arXiv:1310.8214
(2013).

\bibitem{partdata}
J. Beringer  et al. (Particle Data Group), Phys. Rev. D86, 010001 (2012). 

\end{thebibliography}
\end{document}